# Response to Comment on "Non-Polaritonic Effects in Cavity-Modified Photochemistry": On the Importance of Experimental Details

Philip A. Thomas*, William L. Barnes
Department of Physics and Astronomy, University of Exeter, Exeter, EX4 4QL UK
*p.thomas2@exeter.ac.uk

Keywords: Polaritonic chemistry, Strong light-matter coupling, Photochemistry, Cavity polaritons, Photoisomerization

**Abstract:** This note responds to Schwartz and Hutchison's Comment (DOI:10.1002/adma.202404602) on our article (DOI:10.1002/adma.202309393). We think differences have arisen not in the experimental results themselves but in their interpretation: our more extensive experiments allowed us to distinguish between "true positive" and "false positive" results. We identify potential evidence of non-polaritonic effects in Schwartz and Hutchison's own work. We hope our work will encourage others to produce more systematic investigations of strong coupling.

Schwartz and Hutchison's recent Comment [1] largely reiterates the findings of their earlier work [2] and highlights perceived differences between our work [3] and theirs. The authors do not provide evidence that these small differences in experimental design led to a substantial difference in results; indeed, we are not aware of any contradiction between our experimental results and theirs. Instead, we think the difference lies in the interpretation of results.

Our main concern is that Schwartz and Hutchison have missed the central message of our article. We argued that, in their work, they studied too few samples, making it impossible to distinguish between a "true positive" result (i.e. strong coupling modifies the photoisomerisation rate) and a "false positive" result (i.e. a non-polaritonic effect dominates).

We think there is evidence of non-polaritonic effects in their own work [2, 1]: their "on-resonance non-cavity" and "off-resonant non-cavity" samples have very different photoisomerisation rates, even though strong coupling is not present in either sample. In their Comment, Schwartz and Hutchison present calculations (Fig. 5) which appear to confirm that the Ag structures they studied support strongly confined electromagnetic modes in the ultraviolet (the source of the non-polaritonic effects we identified).

One aim of our own study was to highlight the need to quantify non-polaritonic effects before invoking strong coupling to explain experimental results. It may be that one of the leading causes of confusion in the field of polaritonic chemistry is the lack of experiments that are designed with the inherent capacity to enable one to distinguish between "true positive" and "false positive" results. Looking ahead, we hope that new systematic experiments will be carried out from which we may be able to better understand what role(s) strong coupling might play in modifying material properties.